# Rangzen: Anonymously Getting the Word Out in a Blackout

*Adam Lerner*[1]   *Giulia Fanti*[2]   *Yahel Ben-David*[3]   *Jesus Garcia*[3]   *Paul Schmitt*[4]   *Barath Raghavan*[5]

[1]*University of Washington*   [2]*UIUC*   [3]*UC Berkeley*   [4]*UCSB*   [5]*ICSI*

## Abstract

In recent years governments have shown themselves willing to impose blackouts to shut off key communication infrastructure during times of civil strife, and to surveil citizen communications whenever possible. However, it is exactly during such strife that citizens need reliable and anonymous communications the most. In this paper, we present Rangzen, a system for anonymous broadcast messaging during network blackouts. Rangzen is distinctive in both aim and design. Our aim is to provide an anonymous, one-to-many messaging layer that requires only users' smartphones and can withstand network-level attacks. Our design is a delay-tolerant mesh network which deprioritizes adversarial messages by means of a social graph while preserving user anonymity. We built a complete implementation that runs on Android smartphones, present benchmarks of its performance and battery usage, and present simulation results suggesting Rangzen's efficacy at scale.

## 1  Introduction

Over the past decade, the balance of power between citizens and governments has tilted inexorably in the latter's direction. Though there was once a perception of the Internet as a land of radically free communication and activism, it has become clear that like any key infrastructure, the Internet exists under centralized control [65]. While this has proved to be a boon to the average user under placid conditions—large companies are able to deliver efficient and reliable services and governments are able to police networks for criminal activity—in times of unrest, this has proven dangerous to those who would use the Internet as a forum to speak out.

Indeed, during societal unrest, centralized infrastructure can be easily co-opted. In recent years, authorities in Egypt, Iran, and Syria, among others, have shut down their already heavily-surveilled Internet access during times when citizens were questioning those very authorities' political legitimacy [13, 20, 64]. However, it is exactly in such moments that citizens need the ability to communicate without restriction and risk of retribution. In particular, in such a scenario, not only is anonymity important, but also the resilience of the communication system to network-level censorship (at multiple layers).

To address this need we built Rangzen, an anonymous large-scale messaging system robust to network attacks. We built a complete implementation of Rangzen for the Android platform utilizing Bluetooth and WiFi Direct to communicate from pocket to pocket between stock, non-rooted, Android version 4.x, 5.x, and 6.x devices. Our implementation consists of over 17,000 lines of Java and another 4,000 lines of unit tests; our implementation is publicly available [53].

The literature contains a wide range of work on anonymous communication systems. However, most existing work targets point-to-point communication over dedicated infrastructure (e.g., over the Internet, which is unavailable in a blackout context), such as Tor [22]. Another, less-well-known branch of research focuses on point-to-point anonymous communication in the absence of dedicated infrastructure [37]. In addition, some widely-promoted free apps such as FireChat also aim to provide point-to-point communication without infrastructure, but fail to provide anonymity or any mechanisms to withstand attacks by adversaries [29, 55].

Rangzen differs from prior work by focusing on *one-to-many*, *anonymous* communication *without* dedicated infrastructure while providing *resilience* to network level attacks, meeting a need which has arisen with surprising regularity around the world in recent years (§2). To the best of our knowledge, no existing work explores this region of the design space. A core aspect of our work is that we address a key tension between anonymity (which demands that user identities be hidden) and robustness to network-level attacks (which requires some notion of reputation or identity, despite user anonymity). In addition, we focused on designing and building a system that works in practice and meets the needs of real

users—dissidents and activists with whom we have been in contact—in medium-risk contexts.

Two main challenges characterize this context: a) a lack of functioning infrastructure (neither Internet nor cell networks) and b) adversaries intent upon widespread interference. We address the first challenge by broadcasting messages over a delay-tolerant, smartphone-based, mobile mesh network. This architecture enables resilient albeit high-latency communication. We address the second challenge with a decentralized, novel social trust-based prioritization algorithm. Counterintuitively, our prioritization approach allows Rangzen to rank the trustworthiness of messages without knowing their origin or author; this property preserves author anonymity. In this decentralized setting it is not possible to entirely eliminate adversarial messages, as there exists no trusted authority that can authenticate users or messages. However, as a result of Rangzen's ranking algorithm, adversarial messages arrive with low scores that enable easy filtering, and propagate more slowly through the network.

We have worked with activists in countries with widespread censorship and political repression to ensure Rangzen represents a realistic system that addresses challenges real people face in the modern world. As a result, it was key for Rangzen to work on *unmodified* modern mobile devices; a previous generation of mobile mesh protocols using smartphones typically required the ability to craft custom layer-2 frames [47, 66], something only available on rooted phones; ordinary users seldom have rooted phones, so this significantly limits deployability. To work around network stack limitations, we developed a hybrid protocol (§5) that uses a combination of WiFi Direct and Bluetooth and works on unmodified Android devices without needing user interaction.

Our experimental results show Rangzen is practical: devices in proximity for under 10 seconds can find one another, perform cryptographic social-trust prioritization, and exchange several hundred messages (§6). Even in the unrealistic worst case of constant communication, Rangzen uses only about 5.5% of the battery of a Nexus 5 per hour, suggesting that under realistic conditions its energy use would be quite reasonable, especially in a network blackout. Our simulations suggest that if deployed widely, Rangzen would provide useful messaging within a large city (§7). Specifically, we show that Rangzen can reach over 90% of a city-wide userbase within 24 hours of a message's composition while effectively hampering attacks such as jamming and the spread of propaganda. In simulation, an ideally-situated attacker's messages reached 30% fewer users than an ordinary (non-ideal), legitimate user's messages over a period of 48 hours.

## 2 Background and Motivation

The design of a communication system for use during network blackouts and similar settings is a complex undertaking. Social and political dynamics constrain both user desires and adversarial actions. On one hand, it would be naïve to design technology to respond to censorship or blackouts in countries or regions with extreme levels of repression (e.g., North Korea), as far more than a technological solution is required [69]. On the other hand, places with relatively low levels of risk and repression (e.g., Hong Kong) generally do not require specialized solutions such as Rangzen. Thus in this work we target medium-risk regions (e.g., Kazakhstan). In such medium-risk environments, while there exist harsh restrictions on free speech and public protest, especially in Internet-based communication, authorities do not typically resort to open violence or mass arrest.

In this work we have been in contact with dissidents and activists in medium-risk contexts, and this has helped us gain a better understanding of their needs and has informed our design choices. Understanding these constraints is crucial to designing and building a system that is realistic in its aims and limits. In this section we present an abridged analysis of this context and its implications for Rangzen.

### 2.1 The Need for Anonymity

Technology's role in the spread of dissent is complex. In most recent political movements, smartphone- and Internet-based organizing was prevalent across social demographics [7, 11, 25, 33, 46, 74, 81]. Governments responded by leveraging the same communications networks not only to surveil [28] but also to intimidate citizens [77]. Although such technology certainly enables the spread of information, in many cases that very information has helped governments to target activists to an unprecedented degree; this targeting ranges from identifying dissenters online to harassing citizens who possess out-of-the-ordinary communication hardware [26].

**Implication: During periods of social unrest, communication systems should protect individuals from being directly linked to objectionable content.**

### 2.2 The Need for Robustness

The communications blackouts that motivate our work are not accidents, and are due to a desire to stem the tide of public discontent. Nevertheless, the means by which blackouts have been implemented and their degree of totality have varied considerably. Some blackouts have involved BGP route withdrawals [16], while others have been more severe cuts [67]. In addition, total blackouts are made easier for governments to impose by a non-diverse network infrastructure with few providers and/or heavy government control. During these black-outs, governments are likely to try to thwart any temporary workarounds used by the population. A robust network need not ensure connectivity for every single person (as individuals can be targeted) but for the masses.

|  | | **Communication Model** | | |
|---|---|---|---|---|
|  | | One-to-one (secure) | One-to-many (non-secure) | One-to-many (secure) |
| **Infrastructure** | Mesh | Threshold-pivot [37], ALAR [48] | Firechat [55] | *Rangzen* |
|  | Internet | Tor [22], Crowds [63], LAP [35], VPNs [3, 59], HORNET [10], Aqua [42], Tarzan [30], Rome [60], Herd [41], Cashmere [84], Free Haven [21] | Twitter [73], Whisper [78, 79] | DC Nets [8], Dissent [15] |

Table 1: Design space of relevant mesh and anonymous communication systems, with canonical examples. We omit insecure one-to-one communication in this comparison due to its ubiquity.

**Implication: The cause or mechanism of a blackout should not impact the subsequent operation of the system, and the system should resist unilateral, global shut-down or takeover.**

## 2.3 The Paradox of Fast Communication

Recently, political theorists have studied the impermanence of so-called "Twitter revolutions", commenting on the absence of slow, steady community organizing and in-person contact that has led to lasting political movements in the past [72]. The determining factor in the outcomes of such crises is largely an enigma, though the strength of the underlying movements has been a crucial factor [34]. While only history will adjudicate the impact of modern communications tools in these settings, our conversations with activists and the historical record suggests that rapid communication is not crucial.

**Implication: The system need not enable rapid communication, but it should enable trustworthy and robust communication.**

## 3 Related Work

In Table 1 we categorize the design space of related anonymous communications systems. A number of cryptographic, anonymous broadcast protocols exist [6, 9, 15, 45], but it is unclear whether such protocols are well-suited for large-scale adoption in resource-starved environments. The problem is particularly challenging in a blackout, which prevents the use of online trust mechanisms, such as Bitcoin-based protocols [52]. On the other hand, much of DTN security research has focused on one-to-one communication [4, 24, 27]; our problem is more related to filtering content in a distributed and privacy-preserving manner. Any solution in this setting must be lightweight so as to function during short opportunistic encounters. Many practical and academic anonymity systems attempt to resolve this tension using pseudonymous reputation systems [37, 68], but using pseudonyms increases susceptibility to side-channel correlation attacks [19, 54, 62].

Our approach relates to Sybil defense [58, 76, 80, 82] through our use of social graph structure to distinguish users. Similarly, our approach builds on a rich cryptography literature on distributed, privacy-preserving trust [31, 36, 70], and more specifically, private set intersection over social data to achieve privacy and anonymity [43, 44, 71]; we explore our application of this in the next section.

## 4 Architecture and Protocol

Each aspect of Rangzen's design is based on the principles we discussed in §2. The prevalence of smartphones and their use in community organizing enables the basic architecture: a mesh composed entirely of smartphones. The communication model is delay tolerant (§2.3).[1] Rangzen provides anonymity (§2.1) to authors to enable free speech without persecution, and it sidesteps attacks based on pseudonymity by providing full anonymity instead. To provide trust in messages and to suppress propaganda and spam, we filter and prioritize based on social trust between user nodes at propagation time, rather than directly between readers and authors. Borrowing from Sybil defense schemes [82], Rangzen assumes that messages are more trustworthy if they reach a reader via a path of trusted nodes. In Rangzen, we determine trust without knowing the identity of the node which forwarded the message. We use a cryptographic PSI-Ca (Private Set Intersection Cardinality) protocol to determine the number of common peers in the (implicit) social graph shared by communicating nodes; the more friends in common, the more trust assigned to the conveyed messages.

---
[1] Real-time networks over meshes are difficult and potentially fragile, while our delay tolerant system is both robust and is sufficiently fast to enable community organization (messages propagating in minutes to hours, depending on user density and mobility).

## 4.1 Threat Model

Our adversary is a state-level actor capable and willing to disable infrastructure such as cellular networks and ISP networks providing links to the Internet. The adversary's goals are to *disrupt communication* and to inject false information (i.e., *propaganda*). Such an adversary shares similarities to those considered in Sybil defense work—although the adversary may be physically distributed and possess significant technological and financial resources, its weakness is an inability to socially infiltrate its enemies at scale. The adversary's Sybils will nearly always have fewer friendships with honest nodes than the honest nodes have among themselves.[2] This is a common assumption in the Sybil defense literature [76, 82]. A worst-case adversary might violate this assumption by recruiting spies from the general population. However, even the most heavy-handed adversaries in history—such as the Stasi—used only 2% of the citizenry as informants [39]. Moreover, if there are no characteristics distinguishing adversaries from regular citizens, the Sybil detection problem cannot be solved in the first place [23].

**Non-Goals.** We assume that the adversary has the resources to single out individual users and perform severe, targeted, violent or social attacks against them. We do not protect against such targeted attacks. Rangzen aims to make it infeasible for the adversary to *scale* their attacks, making it impossible for them to deceive a large percentage of the population, or to disrupt communications on a large scale. We consider preventing the *scaling* of attacks beyond a small percentage of citizens to be a fundamental win, as have some privacy scholars [32].

While phone-to-phone exchanges are private, message content itself is not private in Rangzen. As with a public system like Twitter, messages are public. By contrast, authorship is confidential and protected. Further, the use of Rangzen is detectable by an attacker. The attacker can participate in the system. Rangzen does not seek to hide the fact that Rangzen is being used. Instead, it relies on decentralization to protect it from attack. Hiding the identities of devices from a local eavesdropper is orthogonal to our goals; while Rangzen does not by default hide the device's identifiers (e.g., Bluetooth and WiFi MAC addresses) as this requires rooting, users with rooted phones can randomize their MAC addresses. In addition, we assume that an attacker cannot easily perform man-in-the-middle attacks at scale against phone exchanges as it would require anticipating and suppressing each pair of phones' physically more proximate communications while injecting false messages during connection establishment. We also do not consider zero-day attacks against Rangzen and PSI-Ca; while these may occur, there is no plausible method for preventing them, and the best remedy is that we have built the system in a modular fashion to enable rapid re-engineering of compromised modules.

More generally, our goal is to enable anonymous and robust communication in situations where the majority of phones are in use by honest users and cellular/Internet infrastructure is not working. As we have learned from our discussions with users in these contexts, real-time communication in these settings is largely needed for disseminating live information to the outside world and satellite is ideal for this; within the blackout region itself, for the majority of users, asynchronous and moderate-speed information dissemination is sufficient.

## 4.2 Design Overview

**Delay-Tolerant Mesh Network.** Rangzen forms a mobile, ad-hoc, delay-tolerant network of smartphones. The network propagates microblogs—broadcasted, public datagrams, much like Tweets—which are passed opportunistically between devices in physical proximity. Messages will spread rapidly through a crowd of people who have Rangzen installed on their phones over the course of tens of seconds or minutes. Messages need not spread in real time; they are stored on devices and forwarded opportunistically when another device running Rangzen is encountered. Users need not actively participate in forwarding messages, as Rangzen runs in the background, periodically searching for nearby devices and sending and receiving messages to and from those devices. Thus messages will also spread over time through a city or region as people move, passing on the street, riding transit, or spending time together.

**Prioritization Based on Mutual Friendships.** Rangzen nodes trust messages forwarded to them by other nodes with whom they share *mutual friends*. Each node stores a priority value in $[0, 1]$ with each message, which is displayed to users and used to order messages in the UI. Low-priority messages are displayed to the user at the bottom of the feed, or not at all. When two users meet and exchange of messages, low trust messages are the last to be forwarded. Low trust messages may be deleted from storage by Rangzen nodes. It is important to note that message spread in Rangzen does not depend upon users sending messages directly (and only) to their friends; instead, two users who have never met before can exchange messages and the trust calculated (but not the communication) depends upon their mutual friends. As such, the communication graph can effectively be more richly connected than the friendship graph.

Nodes determine the trust to put in each message based upon their trust of the device which forwarded the message. Since Rangzen is a fully anonymous system, no authorship information is stored in messages. Trust must be based entirely upon relationships between the nodes, who store and forward the messages. Trust between devices is based upon mutual friendships. When Alice receives messages from Bob, Alice assigns trust to those messages proportional to the number of friends they have in common.

---

[2]In Rangzen these friendships must be made in person, making attacks against friend establishment difficult to scale.

These friendships are real-life relationships, and they must be formed in person through the exchange of secrets between the friends' smartphones. These friend establishments can be made at user's leisure *before* a blackout happens, or during a blackout. Each device uses a secure source of randomness, available via the crypto library, to generate a random 128-bit identifier upon first launch of Rangzen; we assume due to the size of the IDs that they are unique. During friend establishment, Rangzen displays on the screen a QR code containing a hash of the user's Rangzen random ID, which can be scanned by another user to form a friendship. When two devices communicate, they use the PSI-Ca protocol, learning the number of friends they have in common without revealing who their friends are, or even which friends are shared. Each node then assigns trust to messages received in the exchange based on the number of common friends.

**Suppressing Propaganda via the Friend Graph.** Friendships in Rangzen can be viewed as forming a *trust graph* in which each user is a node, and graph edges represent real-life trust relationships between people. We leverage this trust graph to suppress the adversary's propaganda. Since the adversary cannot form friendships with real users on a widespread basis (§4.1), propaganda sent by adversarial nodes will reach genuine users through message exchanges including very few common friends. Thus Rangzen nodes will assign lower degrees of trust to adversarial messages, compared to messages from genuine users. Propaganda messages will tend to be dropped automatically or hidden from user view, reducing their influence on the communications happening through Rangzen.

### 4.3 Peer Exchange Protocol

Rangzen's mesh communication is achieved via lightweight, pairwise message exchanges. Here we define the protocol two peers use when they encounter one another to propagate messages, intersect friend sets, and finally how each uses the information it has received to quantify social trust and discriminate between messages.

**Lightweight Protocol.** The Rangzen protocol is lightweight and simple. Peers establish an encrypted phone-to-phone communication channel; we describe in detail in §5 how we overcome phone limitations to establish this channel. Once the channel is established, we perform a friend set intersection and message exchange in each direction in sequence, encoding all messages using Google protocol buffers to ensure correct parsing. We abort and discard the exchange if any errors occur. We implement the one-round PSI-Ca protocol of Cristofaro *et al.* [17], which is ideal because it only relies upon standard assumptions and Diffie-Hellman group operations that are available in standard Android crypto libraries. Our implementation of this protocol is about 400 lines of Java. This enables two nodes to compute the cardinality of the mutual friend set, but not the identity of those friends. The client message includes not only PSI-Ca information but also the set of messages and priorities that the device knows of.

**Social Trust Metric.** We assume that a) people trust and want to see messages from people socially well connected to them, and b) adversarial nodes cannot infiltrate the social graph at scale. To capture these ideas, each pair of nodes computes a social trust score during each opportunistic encounter. We let $T(a,b)$ denote how much $a$ trusts $b$:

$$T(a,b) = \max\left(\frac{F(a) \cap F(b)}{F(a)}, \varepsilon\right)$$

where $F(a)$ denotes the set of $a$'s friends, and $\varepsilon$ is a small positive constant that ensures that ordering is preserved—even if the nodes share no mutual friends.

We expect users to have around 30 trusted friends, and limit the number of friends that can be submitted to any PSI interaction accordingly, since we consider the trust between people with, for example, 30 and 70 common friends to be similar. This restricts numbers of common friends to small integers, reducing the extent to which common friend degree can distinguish unique users.

**Mapping from trust to priority.** After receiving messages from Bob, Alice's node must decide where to insert those messages in her feed. In our implementation, she simply multiplies the priority score of each message according to the sender by her trust of the sender. These mechanisms may be more effective for ensuring message propagation and the effective filtering of propaganda; we also consider the use of a distorted function that assigns more trust to senders with a threshold number of mutual friends. Additionally, to give message authors increased deniability, each sender adds noise to each message's priority score *before* sending it to a new node. We show in §7 that this improves deniability.

More precisely, if $a$ receives a message from $b$ with priority $0 \leq p_o \leq 1$, then $a$ will insert the message into her queue with a priority that is a sigmoidal function of the trust score:

$$Tr_0^1[(p_{b,a}(T(a,b)) \times p_o) + z_a],$$

where $z_a \sim \mathcal{N}(\mu, \sigma^2)$ is additive Gaussian noise used to improve message propagation,[3] $Tr_0^1[x]$ is a threshold forcing $x$ to be in the range $[0,1]$, and

$$p_{b,a}(T(a,b)) = \frac{1}{1+\exp\{-\rho(T(a,b)-\tau)\}}. \quad (1)$$

In essence this means $a$ will trust $b$ fully if the ratio of mutual friends is greater than $\tau$. We used $\rho = 13$ and $\tau = 0.3$ for a sigmoid that transitioned sharply in the range $[0,1]$.[4] If a device runs out of storage, the lowest-priority messages get dropped first.

## 4.4 Authoring and Message Management

Rangzen presents users with an ordinary microblogging messaging interface. The local user writes new messages, which are initialized to have priority 1. If a user particularly likes a message from another node, she can choose to upvote the message. Note that since messages are anonymous, upvoting a message to priority 1 is equivalent to reauthoring it. Users should avoid revealing their identities through message content; unfortunately, this is difficult to prevent with technological solutions alone, and as with all anonymous communication systems, user education is critical. Finally, Rangzen decays the priority of messages over time so that out-of-date content gradually leaves the system. Low priority messages are shown last, and those below a threshold can be hidden; this ensures that even those adversarial messages that do get propagated do not affect normal users.

## 4.5 Security Discussion

An attacker spreading propaganda must do so via the Rangzen protocol because nodes only accept Rangzen protocol messages. A Rangzen node will only store a new message if the message is authored by the node itself or received during a peer encounter. Attackers that do not corrupt the Rangzen software itself must therefore attack the peer encounter protocol to spread messages.

**Unique device identifiers.** An attacker may attempt to identify individual devices and their owners that are using Rangzen. Rooted devices can randomize the device's MAC addresses on each exchange.[5]

**Propaganda spread.** Rangzen nodes reject messages from peers that do not complete PSI-Ca. If the attacker performs PSI-Ca then their success at spreading propaganda (messages with high priority scores) depends upon the ability to form friendships with real users (§4.1). It is difficult for attackers to form friendships with significant numbers of real users, and as a result their messages will be penalized by the trust score calculation when propagated. The bottleneck between real users and attackers in the trust graph has been used in other application domains, such as Bazaar [58], for a similar purpose.

**Attacking friend addition.** Rangzen only adds friend IDs via in-person exchanges, and these IDs are random and private. Thus attackers must either capture devices or socially engineer targets to befriend users. An attacker who learns friend IDs can store them, forming directed edges in the graph. We rely upon standard device security to ensure the safety of friend IDs on phones themselves. We also designed but did not implement a distributed ID revocation protocol. These enhancements are orthogonal to our design and could be implemented in settings that require them.

**Attacking trust computation.** We place very few demands on the PSI protocol, and our choice of algorithm relies only upon standard (Diffie-Hellman) assumptions. Should the PSI algorithm succumb to cryptanalysis in the future, our modular implementation and design enables its easy replacement. Thus Rangzen is safeguarded against future cryptanalytic breakthroughs.

**Chosen-Input Attack.** Adversaries can learn social graph edges only by submitting IDs to the PSI protocol per encounter, posing as a normal user. The adversary must first acquire such an ID, which is only available to a user's real friends, through another type of attack (e.g., device confiscation). If the attacker includes one ID and the intersection is cardinality 1, the attacker learns that their communication partner is friends with that ID. We call this a chosen-input attack on the trust computation. If an adversary can confiscate significant numbers of devices from users in an intact state, unlock those devices, and extract friend IDs, and then use these IDs to perform a chosen-input attack against peers it meets, this would at most allow the adversary to gradually learn the social graph, as we consider in Appendix B. It is unclear that this is the most efficient way for any adversaries to learn social connections between people; they might instead examine online social media.[6] Given that there is no direct defense of this attack (since the Rangzen setting is one in which there is no trusted authority, so there is no way to prevent Sybil attacks), we enable users to rate limit encounters to reduce information leaked through this channel depending on their level of risk tolerance.

**Denial of Service.** Attackers may attempt to launch denial of service attacks by overwhelming the system with messages. Neither storage nor bandwidth can be overwhelmed by such a flood of messages since the prioritization mechanism applies to all such messages, and clients can terminate exchanges that go on too long. We study attackers that jam the airwaves in §7.2.4.

---

[3] This noise parameter helps unpopular nodes spread content by randomly increasing (or decreasing) priority scores, but it also improves author anonymity. We show this in §7.3.

[4] These constants would need to be tuned in a real deployment based on real mobility and friendship patterns.

[5] Physical-layer device-unique characteristics are more difficult to detect but also harder to conceal on unmodified commodity devices.

[6] An out-of-band social-media based attack is always possible in any system that leverages person-to-person connections. However with Rangzen such attacks are harder because ubiquitous surveillance of Rangzen user exchanges is highly unrealistic. Also note that an external social graph cannot be used to forge identities in Rangzen, as Rangzen IDs are random.

## 5 Implementation

We implemented Rangzen as an Android app. In building Rangzen, we had several goals, including that it be easy for users to use correctly, efficient in terms of battery life, and able to propagate messages quickly and at a distance. Unmodified smartphone platforms have limitations that make this a challenge; in this section we discuss some of those difficulties and our engineering efforts to overcome them. To enable peer exchanges without user interaction, we take a novel approach that combines several technologies available on Android phones since OS version 4.0 (Ice Cream Sandwich): WiFi Direct (known as Wifi P2P in Android) and Bluetooth. Our implementation consists of over 17,000 lines of Java. Our test suite additionally contains over 4,000 lines of Java.

In our conversation with activists, it is customary in medium-risk countries that users who do not have access to an app store will side-load the Rangzen app and share the apk with other interested users through phone-to-phone file sharing. We expect the common case to be that activists who are concerned about privacy will side-load the app after downloading it via Tor, while those who are less concerned will install it from an app store.

We ran Rangzen on 7 models of Android devices: A Nexus 5, a first-generation Nexus 7, a second-generation Nexus 7, a Nexus 4, a Samsung Galaxy S4, a Samsung Galaxy S5, and an HTC One X. These devices were running stock Android 5.0, stock Android 4.4.4, stock Android 4.4.2, Cyanogenmod 11, a Verizon build of Android 4.3, and an AT&T build of Android 4.0.3 at various points in the testing process. These older phones are reflective of devices often in use in countries of interest.

### 5.1 Assumptions

We assumed modern but not cutting-edge devices and operating systems for the users of Rangzen, as this reflects the distribution of devices in countries of interest. Our prototype functions on any version of Android which supports WiFi Direct (4.0 and greater). Android 4.0 was released in 2012 and our target versions include over 97% of Android devices operating today [2].

We assume that requiring users to root their devices would be an unacceptable burden; similarly, we aimed to not require user input to enable passive message forwarding. Thus we rejected approaches which burdened the user in any of these ways.

### 5.2 Engineering Challenges

We found that it was difficult to implement mesh networking capabilities in Android. The hardware supports a variety of protocols which in principle offer convenient peer-to-peer capabilities, including Bluetooth, Bluetooth Low Energy (BLE), ordinary WiFi, hotspot mode WiFi, and WiFi Direct. However, the OS limits the ways we can use these technologies. Here we discuss these limitations before we describe our hybrid solution.

**Ad-hoc WiFi Requires Root.** Ad-hoc WiFi, a classic approach for peer-to-peer communication over WiFi enabled devices, requires a rooted phone. We opted not to require the user to root their phone for the sake of deployability.

**Bluetooth Discoverability Requires User Input.** Bluetooth offers the ability to discover other Bluetooth devices, connect to them, and exchange messages. To be discovered, a device must become *discoverable*, a state in which it broadcasts its presence. For security and user experience reasons, the developers of Android chose to require direct user input any time a device wishes to become discoverable. This model would prevent Rangzen from operating without user intervention. While we ended up using Bluetooth for data transfer, we do not use it to discover peers.

**WiFi Direct Data Requires User Input.** WiFi Direct is a peer-to-peer protocol using WiFi chips that enables devices to discover peers, connect, and exchange data. Unlike Bluetooth, Android's WiFi Direct implementation does not require user intervention for devices to discover one another. However, it does require user input to connect and transfer data with a newly discovered peer.[7] Thus while other devices can be discovered over WiFi Direct, a data connection cannot be formed without user input.

**Bluetooth Low Energy Not Universal.** While BLE provides means for interaction-free communication, it is not universally supported; it requires Android 5+ and special support from phone vendors, and as such is only available on a small number of phone models.[8]

### 5.3 Protocol Design

WiFi Direct enables us to discover other devices without user intervention, while Bluetooth allows connections and data transfer without user input. To work around these limitations, we use both stacks in combination to enable both discovery (WiFi Direct) and data transfer (Bluetooth) without user intervention.

---

[7]The reason for this difference between user input in WiFi Direct and Bluetooth is unclear to us. It may be intentional or accidental. From our perspective as developers, differences like these significantly increased development time, since they are undocumented.

[8]We implemented a prototype BLE stack, but do not rely upon it for our experiments.

Devices searching for peers over WiFi Direct send out beacons to other devices. One of the fields of this beacon is a *name* field, and it is settable in software via a hidden API. We set this name to be the *Bluetooth* MAC address of the local device. Thus we use WiFi Direct discovery to communicate a small amount of information—the Bluetooth MAC address—which is required to bootstrap a Bluetooth connection. Clients now can bypass the discovery portion of the Bluetooth stack; no user interaction is required if the Bluetooth MAC address of the remote device is already known. Thus devices can discover each other and communicate without any user interaction and without rooting the device. WiFi Direct has greater range than Bluetooth, but our effective range is limited by Bluetooth since we must be in range for both technologies.

## 6 Microbenchmarks

Table 2 depicts our benchmarks of our implementation. The *Total* row represents the total time for an exchange plus the time to locate a peer. Each other row represents a small experiment we did to measure individual factors that contribute to the full time in an exchange. The *Other* row represents time measured in a full exchange not accounted for by our measurements of individual factors.

### 6.1 Network

All network benchmarks were performed at a 10 meter distance between a Nexus 5 and a Nexus 7.[9] For each measurement we performed at least 100 trials.

**Measuring an entire exchange.** We measured the time from the beginning to end of an exchange between two devices, after peer discovery. The devices were preloaded with 100 messages (140 bytes each) and 30 friends, which resulted in the transmission of approximately 23.5 KB of data in each direction. In Table 2, we add the peer discovery time to these values to form our "Total" row.

**Cryptographic Operations.** We measured the runtime of the computations performed for the PSI-Ca protocol. Initialization of the PSI-Ca protocol takes 350ms on average. This can be done offline, but we have not implemented this optimization. The online portion takes 260ms on average.

**Peer Discovery.** We measured the wallclock time between calling the Android API that starts a peer-finding scan until the time our application located a nearby peer.

---

[9]We found that even using Bluetooth, Rangzen can operate between devices at distances of 40-50 meters; we limited our experiments to 10 meters since we believed that approximated average distances communicating devices were likely to experience.

| Step | Avg | StdDev | Med | 90th% |
|---|---|---|---|---|
| *Peer Discovery* | 1.18 | 0.70 | 0.83 | 2.18 |
| *BT Connect Delay* | 2.30 | 0.80 | 2.19 | 3.29 |
| *BT Latency* | 0.18 | 0.11 | 0.18 | 0.29 |
| *Data Tx* | 1.52 | 0.18 | 1.56 | 1.75 |
| *Crypto* | 0.61 | 0.01 | 0.61 | 0.62 |
| *Other* | 0.82 | — | 0.81 | 0.60 |
| *Total (measured)* | 6.61 | 1.42 | 6.18 | 8.37 |

Table 2: Breakdown of the time spent during a Rangzen exchange between two peers in seconds.

**Delay of Bluetooth Socket Connection.** After peer discovery, the devices involved are aware of each others' presence but must form a Bluetooth RFCOMM connection before transmitting data. We measured the time between requesting such a connection and being informed by the OS that the connection was ready.

**Latency of the Bluetooth Data Channel.** We measured the round-trip time between two devices which were already connected over Bluetooth. We sent an integer nonce (4 bytes) back and forth over the channel a single time, counting the time between the measuring node's transmission and its receipt of the echo.

**Bandwidth of the Peer-to-Peer Channel.** We measured the raw bandwidth of peer-to-peer Bluetooth links, which can be viewed primarily as a limiting factor on the number of broadcast datagrams we can transmit in a single encounter between peers. These speeds were measured over a payload size of 150KB. We measured a median of 15.09 KB/s. The 90th percentile lowest bandwidth was 13.41 KB/s. In Table 2 above, we converted these bandwidths into the amount of time required to send 23.5 KB at that bandwidth. This corresponds to the amount of data sent in our integrated benchmark for 100 messages and 30 friends.

### 6.2 Battery Usage

A key adoption concern is power drain. We measured the additional battery load imposed on a device, as reported by Android's battery manager. These tests were performed on a Nexus 5. We found that Rangzen consumed 5.5% of the device's battery per hour when communicating nearly continuously (by initiating communication with a nearby device every 10 seconds). This represents the worst case for battery usage, since it involves constant communication. We believe that this is reasonable battery drain given that during a blackout there are few other means of communication, and thus Rangzen is a more valuable app in those circumstances.

## 6.3 Discussion

Bluetooth's bandwidth constraint and connection delay account for about 1/3rd each of the duration of an exchange.[10] Nevertheless, nodes can discover peers and communicate hundreds of messages in opportunistic encounters of less than 10 seconds. As such, even passersby on a street will be in range long enough to permit an exchange.

## 7 Simulations and Analytical Results

Our experiments indicate that Rangzen can disseminate messages quickly at the device level. In this section, we evaluate the anonymity and message-spreading properties of Rangzen at the network level. Rangzen's anonymity and reliability depends on large-population statistics; we conducted several tests with dozens of users over several weeks, but such small-scale experiments with real subjects are not indicative of performance at scale. Therefore, we have simulated Rangzen operating at city-scale over real mobility traces. We also derive anonymity properties theoretically, and evaluate our expressions for the datasets considered. Due to the lack of large public datasets containing both social and mobility data, we have used a large-scale mobility dataset and imposed a social graph using known methods.

For simulating message spread, we used real-world datasets, including mobility traces (EPFL Cabspotting [57], St. Andrews Locshare [5], University of Milano PMTR [51], and Technicolor SIGCOMM [56]) and social graphs (two subgraphs of the Facebook social graph [50,75]). We also tested our algorithms on datasets of mobility *and* social connections [5,12], which in principle is what we want, but we found the datasets to be too sparse in time (Gowalla) and space (St. Andrews) for effective evaluation, though our results with them were substantially similar to those we report here.

Our simulator consists of 2100 lines of Java code built upon MASON [49], a discrete-event multiagent simulation library. Our simulator accepts social network graphs or can generate scale-free random social graphs as needed [1]. The simulator supports various mobility datasets. It replays agent locations over time and agents within 20 m[11] are made to encounter one another with some small probability (our simulations use 0.05). This is meant to simulate unreliable message exchanges for worst-case evaluations.

Nodes can also be adversarial, which causes them to perform physical/MAC layer attacks. We model these attacks in a worst-case analysis by assuming that all nodes within range of the attacking node are unable to communicate at all. We do not allow honest nodes to upvote messages, to ensure that simulation results are lower bounds on message propagation speed.

## 7.1 Summary of Results

Our results indicate that Rangzen could, despite network attacks, continue to deliver predominantly legitimate messages during an Internet blackout while protecting the anonymity of message authors.

**Message propagation.** In simulation, Rangzen delivered messages from honest nodes to over 80 percent of the population within 24-48 hours, depending on the prioritization noise parameters (Figure 1). Figure 3 indicates that messages from *individual* popular nodes may spread up to 33 percent more than those from an entire adversarial coalition of nodes, and those adversarial messages have low priority enabling end-device filtering.

**Robustness of the network.** Figure 4 indicates that Rangzen is robust to jamming attacks even when 10 percent of the population is an attacker using jammers with ranges up to 1.3 km. We believe such an attack to be beyond the capabilities of a likely adversary. At the protocol level, Figure 3 indicates that coalitions of adversarial nodes cannot dominate network resources as they have few friends. A coalition of 6 adversarial nodes in a network of 400 nodes performed only marginally better on average than individual honest nodes selected uniformly at random.

**Protection for users.**
*Authorship deniability.* Users can deny authorship of any message with non-negligible probability (§7.3).
*Device Capture.* If an adversary captures $b$'s device, $b$'s friend IDs are password-protected. Without input from $b$, an adversary can only learn mutual friends via the chosen-input PSI-Ca attack. Even with $b$'s password, friend IDs are not stored—only hashes are.
*Trust Graph Extraction.* A resource-limited adversary cannot learn a significant portion of the trust graph. This effect can be amplified by randomly adding and deleting friends in PSI interactions, and limiting the maximum number of friends that can be fed to the PSI-Ca protocol.

## 7.2 Message Propagation

Our metrics of success for message propagation are a) the time required for a message to reach 90 percent of the honest population, and b) the fraction of honest nodes that have received a message by a given time. These metrics are chosen for use cases like protest organization, in which mobilization depends on a large portion of the population cooperating. All plots are averaged over 40 runs. We use epidemic propagation over infinite-storage devices as an upper bound on the spread rate since a mesh DTN cannot disseminate content faster than flooding if storage is unconstrained. Since mobility is harder to model than social relations, we used the Cabspotting mobility dataset [57] with a randomly-generated Albert-Barabasi social graph [1]. The Albert-Barabasi generative model lets us create arbitrarily-sized social networks, and it displays common properties of social networks like high clustering-coefficient, power-law degree distribution, and short path lengths between nodes.[12]

---

[10]We found that BLE provides no better performance.
[11]Our prototype is able to communicate at ranges up to 40-50m, and sometimes farther.

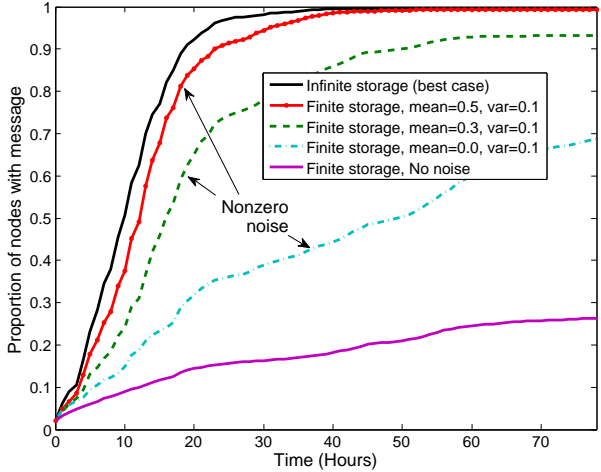
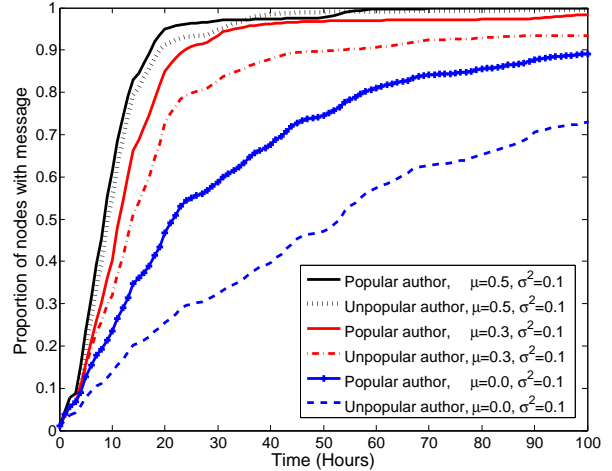

Figure 1: Impact of the Rangzen protocol on legitimate message propagation without an adversary. The additive Gaussian noise in priority scores clearly improves propagation, but may hamper the system's ability to filter out adversarial messages.

Figure 2: Popular nodes can spread messages faster than unpopular nodes. This effect is more pronounced when nodes add less noise prior to transfers (e.g. lower $\mu$). We expect adversarial nodes to be unpopular.

### 7.2.1 Propagation without an adversary

Figure 1 shows the propagation of legitimate messages with no adversary. The curves represent different distributions of the noise parameter $z_i$ in our trust metric. Figure 1 suggests that even using random social graphs, *Rangzen can reach at least 80 percent of the population within 24 hours and 90 percent of the population 20 hours after infinite-storage epidemic routing does so.*

### 7.2.2 Propagation with a passive adversary

Next, we demonstrate the performance of Rangzen under a passive adversary, which deploys devices that follow the Rangzen protocol, but may also disseminate their own content. Distinct groups of friends may wish to emphasize their own content internally without directly attacking others' communications. A node with few connections to a social graph cluster can therefore be considered a passive adversary; its goal is not explicitly to hinder message propagation within the cluster, but messages from more popular nodes in the cluster should be prioritized. This is not a key part of our threat model, but it relates nonetheless to reducing spam in broadcast networks. Figure 2 shows the effects of node popularity on propagation speed. Here, (un)popular nodes were selected randomly from the 5 percent worst- or best-connected nodes in the social network. The figure shows that *messages from popular nodes reach 90 percent of the population as much as 40 hours earlier than messages from unpopular nodes, for certain noise levels.* This model of communication is consistent with natural human communication patterns, which tend to favor people with more social connections.

### 7.2.3 Propaganda Message Spread

Next we consider an active adversary that controls a node coalition. Adversaries have few friends but they can share friend IDs. The adversarial coalition spreads only its own messages. It can create Sybils, but this is of limited use since Sybils do not help befriend honest nodes. We used noise parameters $\mu = 0.0$ and $\sigma^2 = 0.1$ with a community of 400 nodes, using almost the entire dataset.

Figure 3 illustrates the propagation time of messages originating from popular, average, and unpopular honest nodes, as well as the adversarial coalition; we assume that 1.5 percent of the population belongs to the adversarial coalition.[13] The figure shows that *the adversarial coalition can spread messages a little bit better than average nodes, but at least 30 percent worse than* individual *popular nodes*. At very small scales (50 nodes), we observed that average and *unpopular* honest nodes actually performed *better* than the adversarial coalition for the first 48 hours. This suggests that Rangzen can be used by anyone within tighter social circles, though one should be well-connected to communicate at a large scale, just as is often the case in other social networks.

---

[12]However, it does not capture other properties of social graphs such as community development. Also, true social graphs are typically somewhat correlated with mobility patterns.

[13]At its height, the Stasi employed 0.6% of the East German population as agents and another 0.9% of the population consisted of "informal collaborators" [39].

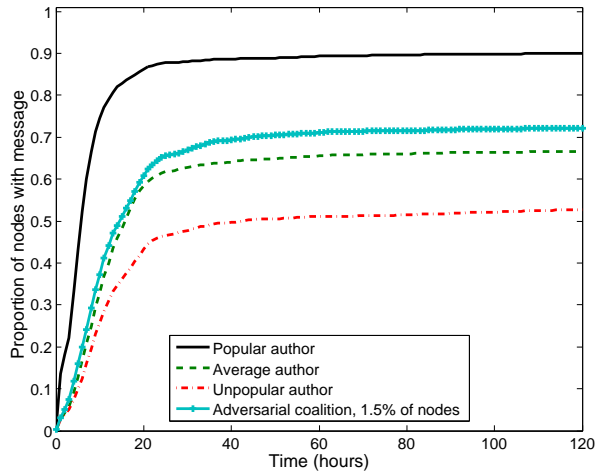

Figure 3: Adversary propaganda spread.

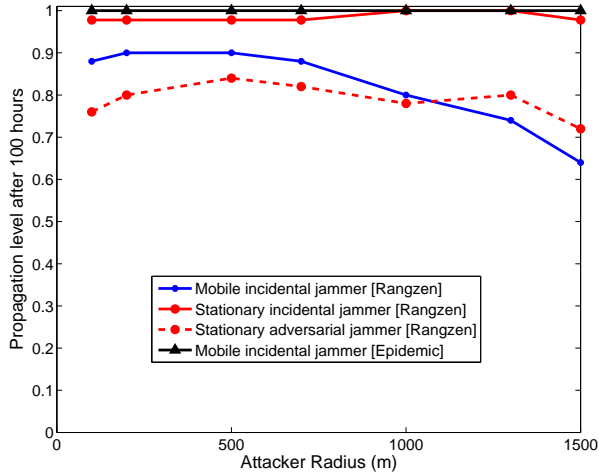

Figure 4: Propagation impact of physical/MAC attacks.

If an adversary were to corrupt popular nodes it could infiltrate a given social circle. Even offline, this is impossible to prevent. Here we must rely on nodes to gradually unfriend corrupted nodes. Similarly, if the adversary corrupts a significant fraction of the population, Rangzen cannot defend against it. We believe that this will be true of any decentralized, mobile-mesh-based solution.

### 7.2.4 Physical Jamming

For a worst-case estimate of physical or MAC-layer attack effects, we first consider a physical-layer attacker (e.g., a jammer). We model it as a point source of omnidirectional radiation in one of the WiFi frequency bands (20 MHz bands at either 2.4 GHz or 5 GHz), as a best-case for the attacker. We assume the attacker targets WiFi Direct rather than Bluetooth, as Bluetooth employs adaptive frequency hopping and channel assessment to avoid interference. To be conservative, we assume the signal follows the path loss formula:

$$P_R = P_T \left( \frac{c}{4\pi d f} \right)^2$$

where $c$ is the speed of light, $f$ is the signal frequency in Hz, $d$ is the distance traversed, and $P_T$ and $P_R$ are the transmitted and received power, respectively (we assume equal antenna gains). We ignore factors like diffraction and absorption, which would significantly weaken a jamming adversary. It is the inverse square factor of path loss that hampers jamming at scale. We estimate the transmit power of a smartphone to be 251 mW (corresponding to average output power over the 5.4 GHz band), and we estimate the maximum output power of a stationary jammer to be 20 W in the same WiFi band (based on military-grade commercial jammers).[14] Under these assumptions, a jammer would need to be within 180 m of the receiver, with line-of-sight, to jam transmissions between nodes 20 m apart. This 180 m attack range is in line with advertised ranges of commercial jammers.

### 7.2.5 Jamming Simulation

We simulate message propagation in a jamming attack scenario. For simplicity, we model a "perfect" jammer (i.e., node pairs located within an attack radius are unable to propagate messages no matter their distance from the attacker). To be conservative we consider jammers with an order of magnitude greater range than we estimated for commercial jammers above, as an attacker might extend this range with MAC-layer attacks (e.g., by sending messages that cause other nodes to not transmit).

We consider mobile and stationary adversaries, both optimally and non-optimally placed. We model mobile, non-optimal adversaries as nodes in the mobility trace. Stationary non-optimal attackers are placed uniformly within the simulation area. We used a simulated annealing algorithm to place optimal, stationary adversaries [38]. An optimal *mobile* attacker would have to know the entire population's location at every instant in time (without the benefit of cell-based location tracking), and solve an NP-hard problem [38]; regardless, we do not believe such adversaries pose a greater risk than attackers traversing popular routes regularly.

Figure 4 shows the impact of such geography-based jamming attacks on message propagation; we find that *even when physical/MAC-layer attackers have omnidirectional ranges up to 1000 m, the system propagates at least 80 percent as well as it does in a best-case non-*

---

[14]Although many military-grade products advertise high overall power output, the bandwidth of such products is typically high and non-configurable; 20 W in the WiFi band is typical.

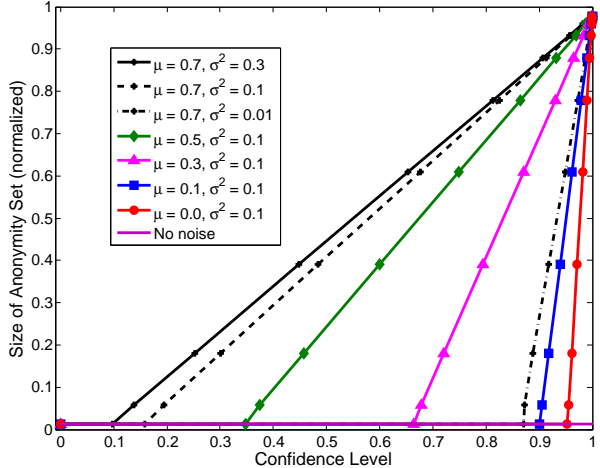

Figure 5: Anonymity set size (fraction of nodes) as a function of the estimator's confidence level. A point $(0.5, 0.2)$ indicates the smallest set of nodes including the author with probability 0.5 contains at least 20% of the nodes.

*jamming scenario*. While such an attack is unlikely, this highlights Rangzen's robustness to localized attacks.

### 7.3 Analytical Privacy Results

Next we evaluate Rangzen's anonymity properties and its resistance to message author identification and social trust graph extraction.

If the adversary receives a high-priority message from an honest node, Rangzen should enable the sender to plausibly deny authorship. For this analysis, we derive a distribution for the *anonymity set*, which is the set of nodes that could have plausibly authored a particular message. Specifically, we estimate how many hops a message took since inception, and then estimate how many nodes are that many hops away for a fixed confidence level. Recall that random noise is added to message priority scores before each transmission. This noise enlarges the anonymity set.

We compute the pmf of the number of hops a message traversed before reaching a target node, given the priority score seen. Suppose node $A$ receives a message from $B$. Let $N$ be the number of hops the message traversed *before* reaching $B$. $S \in [0,1]$ denotes the priority with which $A$ receives the message from $B$ (before considering their mutual friends). $\Omega$ denotes the event that the message is observed by a randomly-selected node (in this case, $A$). For a worst-case analysis, assume that $A$ receives the message with priority $S = 1$. We want $P(N = n | \Omega, S = 1) = P(S = 1 | N = n, \Omega) \cdot P(N = n | \Omega)$.

We present our modeling of $P(N = n | \Omega)$ and $P(S = 1 | \Omega, N = n)$ in greater detail in Appendix A. Using these models, we estimate $P(N = n | \Omega, S = 1)$ as a function of $n$. Combining this with mobility data, we numerically estimate the anonymity set size for a given trace.

Figure 5 shows the size of the author's anonymity set as a function of the estimator's confidence level—the probability that the true author is in the anonymity set—for the SIGCOMM dataset [56]. Using $\mu = 0.3$ and $\sigma^2 = 0.1$, the 90% confidence anonymity set contains 80% of network nodes. More noise significantly increases the anonymity set size. Anonymity is necessarily data-dependent because it is always possible to construct pathological mobility and social trust models in which the adversary can easily identify the source of messages. The correct noise parameters should be selected empirically to balance anonymity with message propagation.

We also consider an adversary who aims to learn global information about the Rangzen trust graph, such as which pairs of nodes are friends. This information can lead to deanonymization through correlation with other social graphs (e.g., Facebook, Twitter) [54]. In Appendix B, we show analytically that due to Rangzen's node ID protection, this is a difficult attack to scale.

## 8 Conclusion

Since the advent of the Internet and the rise of democratized communication there has been a tension between the communication wants and needs of the many and the prerogatives of the few in control of the means of communication. Our aim has been to evade this tension by designing and building a robust, anonymous communication substrate to evade the shutdown of communications infrastructure. We did this by designing a lightweight, anonymous communications protocol; by implementing that protocol in Android; and by examining the behavior of the protocol and of our implementation in a series of benchmarks and simulations, showing that Rangzen is practical and robust at scale. How this tension evolves remains to be seen. An arms race naturally follows the use of circumvention technology like Rangzen. We believe that Rangzen provides both a useful means of communication that is difficult to shut down or co-opt and provides sufficient protection to the average user to prevent retribution by an adversarial government.

### Acknowledgements

We thank Ron Steinherz and Liran Cohen for their contributions to the design and implementation of the user interaction of Rangzen. We also thank the activists and scholars working on anonymity systems we spoke with for their insights.

## A  Anonymity Set Details

A message's priority score $S$ depends on the number of hops the message took. In particular, we can define the received priority after $N = n$ hops ($S_n$) recursively as follows:

$$S_n = p_n \cdot S_{n-1} + z_n$$
$$S_0 = 1$$

where $z_i$ is the noise added by the $i$th node, and $p_i$ is the priority score at the $i$th node. $z_i$'s distribution is designed, but priority scores $p_i$ depend on graph and mobility properties.

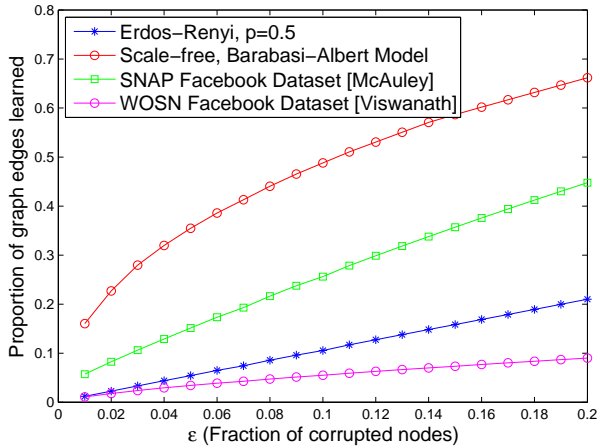

Figure 6: Proportion of graph edges learned by the adversary ($d_\varepsilon$) as a function of $\varepsilon$ (proportion of corrupted nodes).

This priority depends on $p_{i,j}$, the scaling factor when messages pass from $j$ to $i$, as defined in Equation 1. Empirical evidence shows that degree distributions in social networks obey a power law [14]. We found that in the Facebook WOSN dataset [75], mutual degree distributions also obey a power law, but the *ratio $p_{ij}$* across *all node pairs* appears to be better-modeled by a truncated sum of exponentials.[15] This trust metric is not heavy-tailed, so the fraction of nodes with highly overlapping friend sets is very small. This motivates the sigmoid in equation 1, which assigns high trust even if nodes share few friends.

We estimated $P(N = n|\Omega)$ empirically from several datasets. For every pair of nodes in the dataset $(i, j)$, we measured the minimum number of times a message would need to be forwarded before reaching target $j$ from source $i$. This measurement gives an estimated lower bound on how many hops in the (time-varying) connectivity graph separate an arbitrary message from its creator.

## B Deanonymizing the Social Graph

We quantify what the adversary learns about the true social graph through attacks on the private set intersection protocol. As we note in our discussion of non-goals, we do not consider attacks in which the adversary has both confiscated large numbers of user devices and unlocked them (something we are told is uncommon in medium-risk settings though common in high-risk settings) while simultaneously correlating the users' friendship with external social network data sources.

---

[15]Technically, this probability is only defined over rational values, but we approximate the function as having a continuous domain.

Here we assume the adversary knows the nodes $V$ of the social graph. There are many definitions in the literature for graph information content [18]. None of the definitions is clearly superior, so we use the proportion of common edges as a heuristic metric. That is, if the original graph is denoted $G = (V, E)$ and the subgraph is denoted $G_s = (V, E_s)$ with $E_s \subseteq E$, then our similarity metric is $d_\varepsilon(G, G_s) = \frac{|E_s|}{|E|}$.

This metric is closely related to the definition of graph entropy by Rashevsky et al. [61] and was also shown to be strongly correlated with deanonymization success in [54]. Assuming the adversary can corrupt at most fraction $\varepsilon$ of the nodes, we wish to upper bound $d_\varepsilon$ as a function of $\varepsilon$. We show that the adversary will be unable to learn more than 15% of the graph edges by corrupting up to 5% of the nodes, and this can be further limited by artificially adding and removing graph edges during PSI-Ca.

**Static graph.** We first assume the trust graph does not change. As time tends to infinity, we assume the adversary can learn all edges emanating from nodes it has corrupted. This is a worst-case estimate, because it assumes the adversary knows how to align its learned subgraph within the larger trust graph (or a similar social graph from a different domain). In practice, subgraph alignment is difficult.

Figure 6 illustrates the proportion of edges learned as a function of the proportion of nodes corrupted. The SNAP dataset is a Facebook ego-social-circle dataset [50], and the WOSN dataset contains social connections between 55,000 nodes in the Facebook New Orleans network as of 2009 [75]. The figure suggests that as long as the adversary cannot corrupt more than about 5% of nodes, it can learn at most 15% of the social graph. This estimate is worst-case; along with the subgraph alignment issues mentioned earlier, corrupting nodes is difficult, and we expect trust establishment to be less promiscuous in Rangzen than in Facebook.

**Dynamic graph.** Next, we assume that the graph is changing with time. Consider three bins: one with edges learned by the adversary ($L$), one with edges not learned by the adversary ($U$), and one containing edges that are not in the graph ($X$)—i.e., pairs of nodes that are not connected. Each time a new trust relationship is created in this subgraph, another edge is added to the $U$ bin, and each time an edge is deleted (i.e. a user "unfriends" someone) an edge is removed from the $L$ or the $U$ bin. For a worst-case estimate of privacy, we assume the adversary knows when edges are deleted. Edges move from $U$ to $L$ whenever the adversary learns another edge in the graph. Thus we wish to characterize $|L|/|L + U|$. Recall that with a static graph, the adversary could learn at most a small fraction $d_\varepsilon$ of the total edges in the graph.

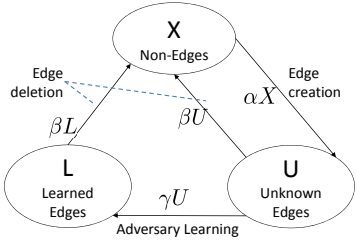

Figure 7: Adversarial learning of a dynamic trust graph.

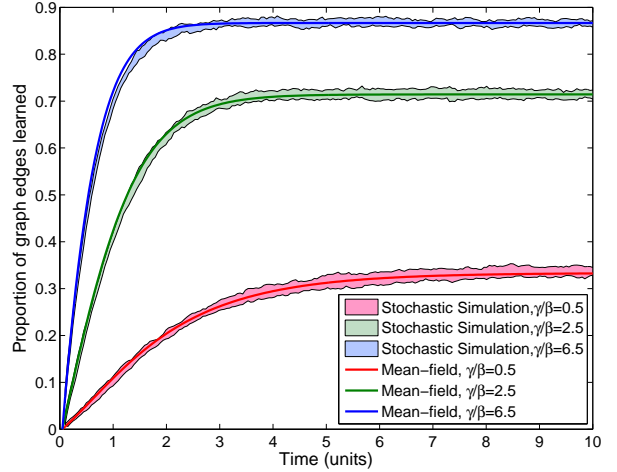

Figure 8: Adversarial graph learning, parameterized by the adversary's learning rate $\gamma$. Asymptotically, the leaked proportion of graph edges depends only on the adversary's learning rate $\gamma$ and the network-wide edge deletion rate $\beta$. We can increase privacy by systematically omitting trust graph edges.

As such, our dynamic model operates within a restricted space of nodes and edges. For instance, if the adversary corrupts 5% of network nodes, then $N_E = X + L + U$ equals the number of edges possible between the corrupted 5% of nodes and the rest of the network. Any equilibrium value of $d_\varepsilon$ in our dynamic model should therefore be multiplied by the results for the static graph.

Our underlying model for this system is a continuous-time Markov chain with Poisson events. The state space of this chain grows exponentially in the number of total possible edges ($N_E$), so we use a mean-field approximation, much like [83]. Figure 7 illustrates our model of the system. $\alpha X$ is the rate of edge creation, $\beta(U+L)$ the rate of edge deletion, and $\gamma U$ the rate at which the adversary learns new edges.

We know that $X(t) = N_E - L(t) - U(t)$ where $N_E$ describes the number of total possible edges. Letting $V(t) = [L(t)\ U(t)]^T$, we have a nonhomogeneous time-invariant linear system:

$$\frac{dV(t)}{dt} = \begin{bmatrix} -\beta & \gamma \\ -\alpha & -(\alpha+\beta+\gamma) \end{bmatrix} V(t) + \begin{bmatrix} 0 \\ \alpha N_E \end{bmatrix} \quad (2)$$

**Observation B.1.** *Let $V(t) = [L(t)\ U(t)]^T$, with dynamics described in Equation 2. Then $\lim_{t \to \infty} \frac{L(t)}{L(t)+U(t)} = \frac{\gamma}{\gamma+\beta}$.*

*Proof.* (Sketch) It is straightforward to show that dynamical system (2) is internally stable, with exact solution

$$\begin{bmatrix} L(t) \\ U(t) \end{bmatrix} = \begin{bmatrix} \frac{\alpha\gamma N(\alpha-\gamma+(\beta+\gamma)e^{-(\alpha+\beta)t}-(\alpha+\beta)e^{-(\beta+\gamma)t})}{(\alpha-\gamma)(\alpha+\beta)(\beta+\gamma)} \\ \frac{\alpha N(\beta(\alpha-\gamma)+-\alpha(\beta+\gamma)e^{-(\alpha+\beta)t}\gamma(\alpha+\beta)e^{-(\beta+\gamma)t})}{(\alpha-\gamma)(\alpha+\beta)(\beta+\gamma)} \end{bmatrix} \quad (3)$$

We then consider $\frac{L(t)}{U(t)+L(t)}$. Since the exponential terms in (3) tend asymptotically to 0, the ratio of interest converges precisely to $\gamma/(\gamma+\beta)$. □

Figure 8 illustrates our analytic results compared to simulated results. The colored bands are inter-quartile ranges over 40 trials. These results affirm our mean-field approximation, not the assumption of constant-rate learning and social graph alterations. However, our model does capture the observation that social graph properties stabilize globally over time, despite continuing to change locally [40].

This result says two things: 1) if no edges are deleted, the adversary eventually learns the entire graph, and 2) asymptotic behavior is independent of the edge creation rate. Over a long time scale, we cannot rely on natural social graph growth to limit the adversary's knowledge. Instead, we should artificially simulate the deletion of edges by (for instance) including random subsets of users' friend sets in each PSI-Ca computation, something we implement in Rangzen.